\documentclass[aip,%
               preprint,%
               citeautoscript,%
               bibnotes,%
               amsfonts,%
               amssymb,%
               amsmath,%
               a4paper,%
               titlepage,%
               fleqn,%
               endfloats*%
               ]{revtex4-1}

\usepackage{graphicx}
\usepackage{xspace}
\usepackage{color}
\usepackage{algorithm}
\usepackage[noend]{algpseudocode}
\usepackage{caption,subcaption}

\begin{document}

\newcommand{\hoch}[1]{$^{\text{#1}}$}
\newcommand{\tief}[1]{$_{\text{#1}}$}
\def\MP{\text{MP2}\xspace}
\def\emp{E_{\text{MP2}}\xspace}
\def\ej{e_{\text{J}}\xspace}
\def\ek{e_{\text{K}}\xspace}
\def\bfL{\mathbf{L}\xspace}
\def\bfB{\mathbf{B}\xspace}
\def\bfP{\mathbf{P}\xspace}
\def\bfS{\mathbf{S}\xspace}
\def\bfV{\mathbf{V}\xspace}
\def\bfa{\mathbf{a}\xspace}
\def\bfb{\mathbf{b}\xspace}
\def\bfc{\mathbf{c}\xspace}
\def\bfd{\mathbf{d}\xspace}
\def\bfr{\mathbf{r}\xspace}
\def\to{\text{o}\xspace}
\def\tv{\text{v}\xspace}
\def\gm{\mu}
\def\gn{\nu}
\def\gk{\kappa}
\def\gl{\lambda}
\def\po{\underline{P}}
\def\pv{\bar{P}}
\def\lo{\underline{L}}
\def\lv{\bar{L}}
\def\re#1{\text{Re}\left(#1\right)}
\def\im#1{\text{Im}\left(#1\right)}
\def\ogm{\underline{\mu}}
\def\ogn{\underline{\nu}}
\def\ogk{\underline{\kappa}}
\def\ogl{\underline{\lambda}}
\def\ogs{\underline{\sigma}}
\def\vgm{\bar{\mu}}
\def\vgn{\bar{\nu}}
\def\vgk{\bar{\kappa}}
\def\vgl{\bar{\lambda}}
\def\vgs{\bar{\sigma}}

\title{Relativistic Cholesky-decomposed density matrix MP2}

\author{Benjamin Helmich-Paris}
\email{helmichparis@kofo.mpg.de}
\affiliation{Max-Planck-Institut f{\"u}r Kohlenforschung,
Kaiser-Wilhelm-Platz 1,
D-45470 M{\"u}lheim an der Ruhr}

\affiliation{Section of Theoretical Chemistry, Vrije Universiteit Amsterdam, 
             De Boelelaan 1083, 1081 HV Amsterdam, The Netherlands}

\author{Michal Repisky}
\affiliation{Hylleraas Centre for Quantum Molecular Sciences, Department of Chemistry, UiT The Arctic University of Norway, 
             N-9037 Trom{\o} Norway}

\author{Lucas Visscher}
\affiliation{Section of Theoretical Chemistry, Vrije Universiteit Amsterdam, 
             De Boelelaan 1083, 1081 HV Amsterdam, The Netherlands}

\date{\today}

\begin{abstract}
In the present article, we introduce the relativistic Cholesky-decomposed
density (CDD) matrix second-order M{\o}ller--Plesset perturbation theory (MP2) energies.
The working equations are formulated in terms of the usual intermediates
of MP2 when employing the resolution-of-the-identity approximation (RI) for
two-electron integrals. 
Those intermediates are obtained by substituting the occupied and virtual quaternion pseudo-density matrices
of our previously proposed two-component atomic orbital-based MP2 (\textit{J. Chem. Phys.} \textbf{145}, 014107 (2016))
by the corresponding pivoted quaternion Cholesky factors.
While working within the Kramers-restricted formalism, we obtain a formal spin-orbit overhead of 16 and 28
for the Coulomb and exchange contribution to the 2C MP2 correlation energy, respectively,
compared to a non-relativistic (NR) spin-free CDD-MP2 implementation.
This compact quaternion formulation could also be easily explored in any other algorithm to compute the 2C MP2 energy.
The quaternion Cholesky factors become sparse for large molecules and,
with a block-wise screening, block sparse-matrix multiplication
algorithm, we observed an effective quadratic scaling of the total
wall time for heavy-element containing linear molecules with increasing system size.
The total run time for both 1C and 2C calculations
was dominated by the contraction to the exchange energy.
We have also investigated a bulky Te-containing supramolecular complex.
For such bulky, three-dimensionally extended molecules 
the present screening scheme has a much larger prefactor and is less effective. 
\end{abstract}

\maketitle

\section{Introduction} \label{intro}

In recent years, quantum chemistry has made
a substantial step forward in applying accurate
wave function-based methods to large molecules
and solids.
Besides several flavours of fragment-based approaches,\cite{Li2004,Friedrich2007,Ziolkowski2010,Rolik2013,Nagy2017}
in particular the revived pair-natural orbital approach of Neese and co-workers\cite{Neese2009,Neese2009b,Riplinger2013a,Schwilk2017} lead to efficient implementations 
that allowed coupled cluster energy calculation of molecules with hundreds of atoms --- even proteins.\cite{Riplinger2013a,Schwilk2017}

An alternative approach is to formulate the working equations
completely in the atomic orbital (AO) basis by means of
a Laplace transformation of the orbital energy 
denominator.\cite{Almloef1991,Haeser1992,Haeser1993,Ayala1999,Surjan2005,Kobayashi2006}
In the AO basis all intermediates including the integrals
become sparse for extended systems.
By employing screening techniques that account for the
rapid decay of the MP2 energy with respect to inter-electronic distances\cite{Lambrecht2005,Lambrecht2005a,Maurer2012,Maurer2013b}
efficient implementations were presented by Ochsenfeld and
his co-workers that allowed calculations on molecules with
more than 2000 atoms and 20,000 basis functions.\cite{Doser2009,Maurer2013b}
A reformulation in terms of AOs by means of Laplace and related transformations
has also been pursued for properties and other electronic structure methods:
MP2 analytic first-derivatives,\cite{Schweizer2008,Vogler2017} MP2 nuclear magnetic shieldings,\cite{Maurer2013}
explicitly correlated MP2-F12 energies,\cite{Hollman2013} periodic MP2 energies,\cite{Ayala2001} 
Dyson correction to quasi-particle energies,\cite{Pino2004} CCSD energies,\cite{Scuseria1999} perturbative triples correction (T) to CCSD,\cite{Constans2000} direct random-phase approximation (dRPA) energies,\cite{Schurkus2016,Luenser2017} and multi-reference second-order perturbation theory (NEVPT2).\cite{Helmich-Paris2017}
However, screening of intermediates in the AO basis 
has a substantial overhead compared to conventional molecular orbital (MO) implementations 
if large basis sets
in combination with diffuse functions are used in the MP2 calculation.
Those extended basis sets are necessary to describe dispersion interactions 
between non-covalently bonded molecules accurately,
which is one of the target applications of MP2 and coupled cluster methods.

To reduce the pre-factor of AO-MP2, a pivoted Cholesky decomposition
of AO density matrices has been proposed by the Ochsenfeld group.
Those Cholesky-decomposed densities (CDD) can preserve the sparsity of the
AO density matrices while reducing the rank,
which is at most the number of active occupied or virtual orbitals.
Consequently, CDD can be considered as generating localized molecular orbitals
for the occupied and virtual orbital space.
With a CDD-based MP2 calculations larger basis sets with
diffuse functions should be feasible to describe also dispersion
interactions in large supra-molecular complexes accurately.

Recently, the relativistic two-component (2C)
MP2 method based on Kramers-restricted formalism has been reformulated
and implemented in the AO basis by the present authors.
In was shown that the working equations of spin-free
non-relativistic and 2C AO-MP2 differ merely
by their algebra, i.e. real versus quaternion, respectively.
The quaternion formulation results in a maximum reduction of the working equations
if point-group symmetry is not considered.
We could also show that the imaginary parts of the quaternion intermediates,
which represent the spin-orbit (SO) contribution to the correlation energy,
are much smaller in magnitude than the real part of the intermediates.
With an implementation that screens 
both negligible contributions
from spatially well separated orbitals 
and small SO contributions of light elements, 
all-electron 2C-MP2 calculations of large 
heavy-element containing molecules should be feasible.

In the present paper, we reformulate the relativistic 2C AO-MP2
in terms of quaternion Cholesky-decomposed densities to allow also calculations
on larger molecules with larger basis sets.
To reduce the overhead of an AO-MP2 calculation even further,
we approximate the two-electron integrals by the resolution-of-the-identity
approximation\cite{Feyereisen1993,Weigend1997} combined with 
attenuated Coulomb operators\cite{Adamson1999,Jung2005} for
a more compact local auxiliary basis.
We show the performance of the implementation in terms of total wall time,
scaling with respect to the system size, and errors introduced by screening
for linear chains of Te-substituted polyethylene glycol oligomers.
Our block-wise screening protocol seems to be less effective
for bulky molecules.
 
\section{Theory and implementation} \label{theory}

\subsection{Relativistic two-component Laplace-transformed AO-MP2}
In our previous work, we have introduced a formulation of
relativistic 2C MP2 energies within the Kramers-restricted
formalism solely in terms of scalar AO basis functions.\cite{Helmich-Paris2016c}
Our reformulation in the AO basis is based on
the numerical integration of the Laplace transform
of orbital energy denominators,
\begin{align} \label{laplace}
\frac{1}{x} &= \int_0^{\infty} \exp(-x\, t)\, dt
 \approx \sum_{z=1}^{n_z} \omega_{z}\, \exp(-x\, t_{z}) \\
x &= \varepsilon_a - \varepsilon_i + \varepsilon_b - \varepsilon_j
\text{,}
\end{align}
that contain orbital energies $\varepsilon$, in which the occupied orbitals are indexed with $i,j$ and the
virtual molecular orbitals (MO) or spinors are indexed with $a,b$.
In the remainder of the present article, we will follow the Einstein summation convention.
As for non-relativistic (NR) AO-MP2, the Laplace transformation in Eq.\ \eqref{laplace}
allows us to compute the 2C Coulomb and exchange MP2 correlation
energy contributions $\ej^{(z)}$ and $\ek^{(z)}$ at each quadrature point $z$,
\begin{align}
 \emp &= - \sum_{z=1}^{n_z} \left( \ej^{(z)} - \ek^{(z)} \right) \\
 \ej^{(z)} &=  
         2\,   \re{ ( \boldsymbol{ \ogm \vgn } | \gk \gl)^{(z)} } 
\re{ (\gm \gn | \boldsymbol{ \ogk \vgl } )^{(z)} }  \label{ej} \\
 \ek^{(z)} &= 
\phantom{2\, } \re{ (\boldsymbol{\ogm \vgn} | \boldsymbol{ \ogk \vgl } )^{(z)} } 
                ( \gk \gn | \gm \gl ) \label{ek}
\end{align}
from two-electron AO integrals  $(\gm\gn|\gk\gl)$ that were transformed 
by occupied
\begin{align}
\mathbf{\po}
 &= \sqrt[4]{|\omega_z|}\, \mathbf{C}^{\to} \times e^{+{\boldsymbol\varepsilon}^{\to} t_z}  \times ( \mathbf{C}^{\to})^{H}  \label{psdenmo-o} \\
 &= \sqrt[4]{|\omega_z|}\, e^{+t_z\, \mathbf{P} \times \mathbf{F}} \times \mathbf{P}   \label{psdenao-o} \\ 
\mathbf{P} &= \mathbf{C}^{\to} \times (\mathbf{C}^{\to} )^{H}
\end{align}
and virtual quaternion pseudo-density matrices
\begin{align}
\mathbf{\pv}  
 &= \sqrt[4]{|\omega_z|}\, \mathbf{C}^{\tv} \times e^{-{\boldsymbol\varepsilon}^{\tv} t_z}
                         \times  (\mathbf{C}^{\tv})^{H}  \label{psdenmo-v} \\
 &= \sqrt[4]{|\omega_z|}\, e^{-t_z\, \mathbf{Q} \times \mathbf{F}} \times \mathbf{Q} \label{psdenao-v} \\
\mathbf{Q} &= \mathbf{C}^{\tv} \times (\mathbf{C}^{\tv})^{H}
\text{.}
\end{align}
The overlap charge distribution $\Omega_{\gm \gn}$ as it appears in 
two-electron AO integrals is transformed as follows:
\begin{align} \label{ocd}
\boldsymbol{\ogm \vgn} &= \boldsymbol{\Omega}_{\ogm \vgn} = 
\mathbf{\po}_{\gm \gm'} \times \Omega_{\gm'\gn'} \times \mathbf{\pv}_{\gn' \gn}
\end{align}
Note that in the equations above bold symbols indicate quaternions.
The quaternion (pseudo-) density matrices in Eqs.\ \eqref{psdenmo-o} and \eqref{psdenmo-v} are computed from
real orbital energies $\boldsymbol{\varepsilon}$ and quaternion molecular orbitals,
\begin{align}\label{qspinor}
&\mathbf{C} = \mathbf{C}^0 + \check{i}\, \mathbf{C}^1  + \check{j}\, \mathbf{C}^2  + \check{k}\, \mathbf{C}^3 \\
&\mathbf{C}^* = \mathbf{C}^0 - \check{i}\, \mathbf{C}^1  - \check{j}\, \mathbf{C}^2  - \check{k}\, \mathbf{C}^3 \\
&\mathbf{C}^H = (\mathbf{C}^*)^T
\text{,}
\end{align}
which are the solutions of Dirac-Hartree-Fock equations in the
Kramers-restricted formalism with maximum time-reversal symmetry reduction.\cite{Saue1997,Saue1999}
Note that $\to$ and $\tv$ denote the active occupied and, respectively, virtual part
of $\boldsymbol{\varepsilon}$ and $\mathbf{C}$.
Alternatively, the occupied and virtual quaternion pseudo-density
matrices can be computed from the quaternion AO Fock matrix $\mathbf{F}$
and the quaternion occupied $\mathbf{P}$ and virtual $\mathbf{Q}$ 
density matrices, respectively, as given in Eqs.\ \eqref{psdenao-o} and \eqref{psdenao-v}.

We chose $\times$ to indicate a non-commutative 
quaternion multiplication for which
the three imaginary units obey the following multiplication rules:
$\check{i}^2=\check{j}^2=\check{k}^2=\check{i}\check{j}\check{k}=-1$.
For notational convenience, we have also used $\times$ to indicate
products of quatertions with real numbers.
The equations presented in this section are completely equivalent to
spin-free non-relativistic AO-MP2 iff we switch from
quaternion to real algebra, i.e.\  omit all imaginary parts.

\subsection{
Relativistic two-component Laplace-transformed CDD-RI-MP2
}
The rational behind a re-formulation in the AO basis
is that by employing localized objects as AO basis functions
the number of non-negligible  contributions
to the correlation energy should scale linearly
with the system size for large molecules.
In particular for MP2
screening of small contributions should be effective
as $\emp$ decays as $\mathcal{O}(R^{-6})$ with $R$
being the distance between two well separated 
charge distributions.

Nevertheless, one-index transformations of two-electron integrals
by AO density matrices in Eqs.\ \eqref{ocd}
introduce a substantial computational
overhead compared to conventional implementations,
which first reduce the dimension by partial transformations to the occupied scalar 
or spinor MO basis. 
An established approximation to reduce the costs of 
both conventional and local implementations of MP2
is the resolution-of-the-identity (RI) approximation\cite{Feyereisen1993,Weigend1997}
that decomposes the four-center AO integrals by a product of two- and three-index
intermediates:
\begin{align}
 & (\gm\gn|\gk\gl) \approx B_{\gm\gn}^P B_{\gk\gl}^P  \label{ri-1} \\
 & B_{\gm\gn}^P = (\gm\gn | Q) \left[ \mathbf{V}^{-1/2} \right]_{PQ} \label{ri-2} \\
 & V_{PQ} = (P|Q)  \label{jpot}
\end{align}
In Eqs.\ \eqref{ri-1} and \eqref{ri-2} $P$ and $Q$ are real atom-centered
auxiliary basis functions.
The errors introduced by the RI approximation with a
Coulomb metric \eqref{jpot} are usually less than 100~$\mu$E\tief{H} per atom
if auxiliary basis sets are used that were optimized for a given
orbital basis set.\cite{Weigend2002}

\begin{figure}
 \centering
 \scalebox{0.7}{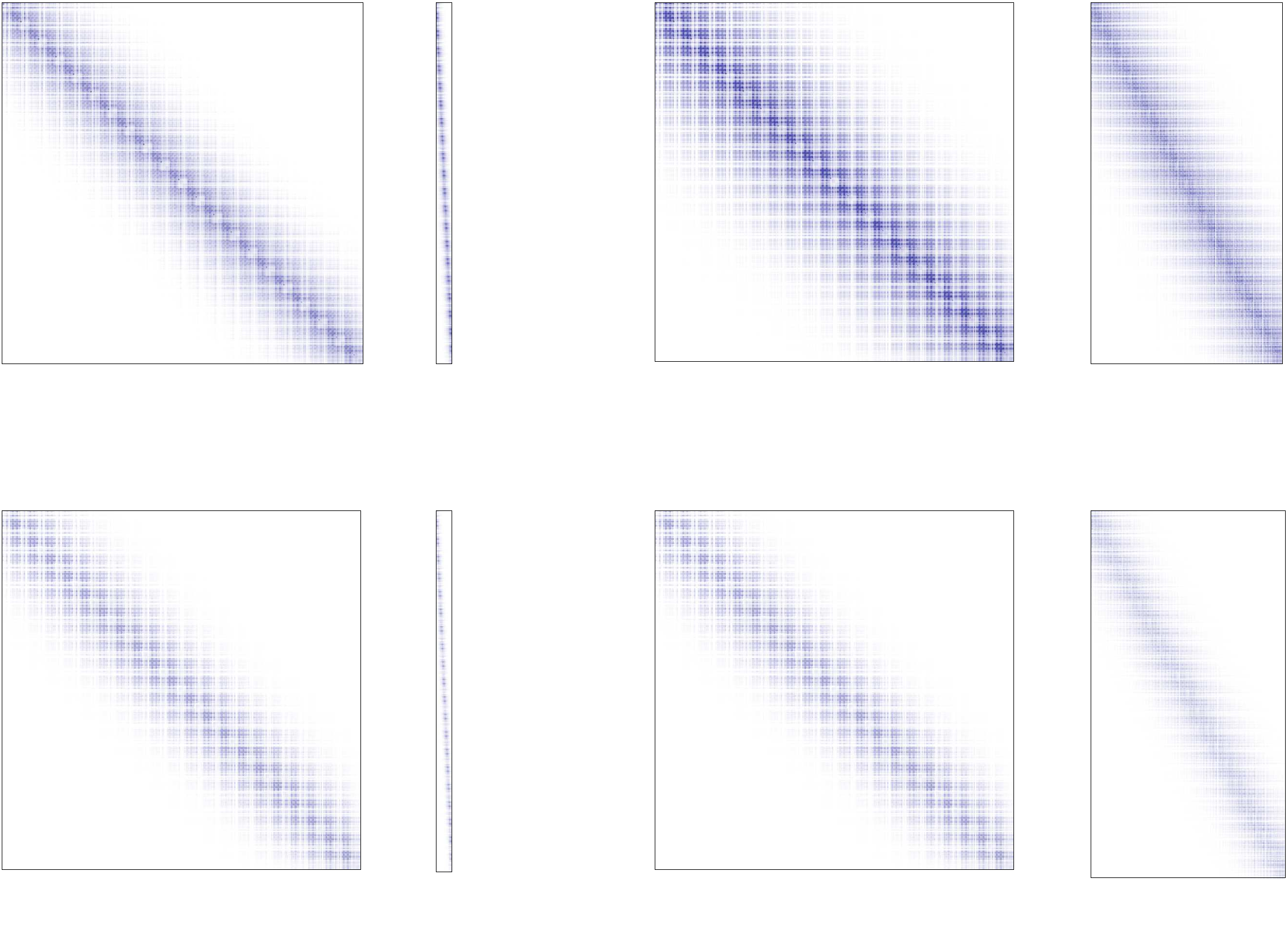}
 \caption{
  Real (0) and first imaginary part (1) of the occupied and virtual
  quaternion pseudo-density matrix and pivoted quaternion Cholesky factors of Te-PEG-20.
}
 \label{fig:te-peg-20-den-cdd}
\end{figure}

The computational costs of the AO-MP2 can be reduced even
further if the pseudo-density matrices
are decomposed into their Cholesky factors.\cite{Zienau2009,Maurer2014b,Luenser2017}
For 2C AO-MP2 the quaternion pseudo-density
matrices should preferably be decomposed by quaternion pivoted Cholesky decomposition (CD)
\begin{align} \label{cdd}
 \bfP &= \bfL \times (\bfL)^H
\end{align}
to preserve the quaternion structure.
In Eq.\ \eqref{cdd} and in the following
we assume that the Cholesky factors have been pivoted already,
which destroys their triangular structure.\cite{Higham2009}
As for the NR case, pivoted quaternion CD
of sparse quaternion (pseudo-)density matrices results in sparse
Cholesky factors $\bfL$, which is illustrated in Fig.\ \ref{fig:te-peg-20-den-cdd}.
To maintain locality as much as possible,\cite{Luenser2017}
the quaternion pseudo-density matrices are transformed 
into an orthogonal AO basis prior to CD:
\begin{align}
& \bfS = \bfL' (\bfL')^T \\
& \tilde{\bfP} = (\bfL')^{-1} \times \bfP \times (\bfL')^{-1} = \tilde{\bfL} \times (\tilde{\bfL})^H \\
& \bfL = \bfL' \times \tilde{\bfL}
\end{align}
We adjust the ordering procedure for the Cholesky
factors of Kussmann et al.\cite{Kussmann2015} for quaternions.
The weighted mean index in Ref.\ \onlinecite{Kussmann2015} is computed
from the norm of a quaternion and
the same column permutations are performed for all four quaternion components.

The 2C-CDD-E\tief{J} is obtained if the quaternion pseudo-density matrices Eqs.\ \eqref{psdenmo-o} and \eqref{psdenmo-v}
are replaced by their corresponding quaternion Cholesky decomposition Eq.\ \eqref{cdd}:
\begin{align}
 e_J &=  \re{ ( \boldsymbol{ \ogm \vgn } | \gk \gl) } 
         \re{ (\gm \gn | \boldsymbol{ \ogk \vgl } ) } \notag \\
     &=  \re{ \bfL_{\gm i} \times \bfB_{ia}^P \times \bfL_{\gn a}^* } B_{\gk \gl}^P \, 
         B_{\gm \gn}^Q \, \re{ \bfL_{\gk j} \times \bfB_{jb}^Q \times \bfL_{\gl b}^* } \notag \\
     &=  \re{ \bfB_{ia}^P \times \bfL_{\gn a}^* \times  \bfL_{\gm i} } B_{\gm \gn}^Q \,
         \re{ \bfB_{jb}^Q \times \bfL_{\gl b}^* \times \bfL_{\gk j}} \, B_{\gk \gl}^P \notag \\
     &=  \re{ (\bfB_{ai}^P)^* \times \bfB_{ai}^Q } \,
         \re{ (\bfB_{bj}^Q)^* \times \bfB_{bj}^P }  \label{ej-bai} \\
     &=  Z_{PQ} \, Z_{PQ} \label{ej-zint}
\end{align}
with
\begin{align}
 \bfB_{ai}^P &= \bfL_{\gm a}^* \times B_{\gm \gn }^P \times \bfL_{\gn_i} \label{bai}
 \text{.}
\end{align}
In Eq.\ \eqref{ej-bai} we exploited 
that multiplication of a real number $a$ and a quaternion $\bfb$ is commutative,
\begin{align}
[a,\bfb]_{\times} = a \times \bfb - \bfb \times a = 0 \label{qcommu}
\text{,}
\end{align}
and that
the real part of multiple quaternion products is 
invariant under cyclic permutations:
\begin{align}
 \re{ \bfa \times \bfb \times \bfc } = 
 \re{ \bfb \times \bfc \times \bfa } = 
 \re{ \bfc \times \bfa \times \bfb } \label{qcycle}
\end{align}

Similarly, we proceed with the exchange energy
\begin{align}
 e_K
&= \re{ (\boldsymbol{\ogm \vgn} | \boldsymbol{ \ogk \vgl } ) } ( \gk \gn | \gm \gl ) \notag \\
&= \re{ \bfL_{\gm i} \times \bfB_{ia}^P \times \bfL_{\gn a}^* \times \bfL_{k j} \times \bfB_{jb}^P \times \bfL_{\gl b}^* } \, B^Q_{\gk \gn} \, B^Q_{\gm \gl} \\
&= \re{ (\bfB_{ai}^P)^* \times \bfB_{aj}^Q \times (\bfB_{bj}^P)^* \times \bfB_{bi}^Q } \label{ek-tmp} 
\end{align}
The non-commutativity of the quaternion multiplication entails us on contracting
either one occupied or one virtual index in a product of two B-intermediate tensors \eqref{bai},
which is not optimal for an implementation that aims for efficiency.
Instead, we would like to contract over the auxiliary basis functions.
This requires an order change in the quaternion multiplication, i.e. swapping the
second and third term of Eq.\ \eqref{ek-tmp}.
As quaternion multiplications are non-commutative, we have to add a commutator
term when computing $\ek$:
\begin{align}
e_K &= \re{ (\bfB_{ai}^P)^* \times \bfB_{aj}^Q \times (\bfB_{bj}^P)^* \times \bfB_{bi}^Q } \notag \\
    &= \re{ (\bfB_{ai}^P)^* \times (\bfB_{bj}^P)^* \times \bfB_{aj}^Q \times \bfB_{bi}^Q } \notag \\
    &- \re{ (\bfB_{ai}^P)^* \times \Big[ (\bfB_{bj}^P)^*,  \bfB_{aj}^Q \Big]_{\times} \times \bfB_{bi}^Q } \label{cdd-ek}
\end{align}
The commutator term leads to 12 additional matrix multiplications and additions, which is given below
in a general form for reasons of notational convenience:
\begin{align}
 \re{ \bfa^* \times [ \bfc^*, \bfb ]_{\times} \times \bfd } = 2 \Big( 
  & a_1 c_1 (b_2 d_2 + b_3 d_3) + a_2 c_2 (b_3 d_3 + b_1 d_1) + a_3 c_3 (b_1 d_1 + b_2 d_2) \notag \\
 +& a_0 c_1 (b_2 d_3 - b_3 d_2) - a_2 c_3 (b_1 d_0 + b_2 d_3) + a_3 c_2 (b_1 d_0 - b_3 d_2) \notag \\
 +& a_0 c_2 (b_3 d_1 - b_1 d_3) - a_3 c_1 (b_2 d_0 + b_3 d_1) + a_1 c_3 (b_2 d_0 - b_1 d_3) \notag \\
 +& a_0 c_3 (b_1 d_2 - b_2 d_1) - a_1 c_2 (b_3 d_0 + b_1 d_2) + a_2 c_1 (b_3 d_0 - b_2 d_1)
\Big)
\end{align}

Our 2C CDD-MP2 formulation uses a pivoted quaternion CD of both occupied and
virtual pseudo-density matrices.
Alternatively, one could decompose only the occupied pseudo-density matrices
to benefit from a rank reduction by using occupied Cholesky factors and
the sparsity of the virtual pseudo-density matrices.
Both approaches were advocated by Ochsenfeld and
his co-workers and lead to successful nearly linearly scaling non-relativistic
implementations for large system.\cite{Maurer2014,Luenser2017}
The correctness of Eqs.\ \eqref{ej-bai} and \eqref{cdd-ek} was confirmed by comparing
the 2C-MP2 correlation energies of our CDD-based implementation
with those from the Kramers-unrestricted RI-MP2 implementation\cite{Bischoff2010}
in Turbomole.

\subsection{
Implementation details
}

A reduction of the computational work is achieved by screening 
Cholesky factors and the three-index integrals when transforming the three-index integrals
to the local CDD, i.e. pseudo-MO basis.
We follow the recipe developed by Kussmann and Ochsenfeld\cite{Kussmann2007b} in which
the Cholesky factors and three-index AO-integrals for a given auxiliary basis function shell
are divided into blocks of a given target block size.
Only those blocks that have a Frobenius norm larger than a user given sparse-matrix threshold
$T_{\text{sparse}}$
are processed further and stored in a block compressed sparse row format (BCSR) with
variable block size.
Before three-index integrals and Cholesky factors are screened and packed
in blocked sparse matrices, the atoms of the molecule are re-ordered to minimize
the band width of a connectivity matrix by the reverse Cuthill-McKee algorithm.\cite{Cuthill1969,Kussmann2007b}
When multiplying two sparse matrices we filter negligible elements on-the-fly.\cite{Borstnik2014}
After the multiplication, we inspect the product
blocks and discard those for which the Frobenius norm 
is smaller than $T_{\text{sparse}}$.
 
In addition to pre-screening of small intermediates, significant computational savings can be made
if the Coulomb operator of the three- and two-index integrals
is attenuated by the complementary error function\cite{Jung2005,Jung2007,Luenser2017},
\begin{align}
  & (\gm\gn|\gk\gl) \approx (\gm\gn|P)_{\omega} \, (\tilde{V})_{PQ} \, (Q|\gk\gl)_{\omega} \\
  & \tilde{\mathbf{V}} = \mathbf{V}_{\omega}^{-1}\, \mathbf{V} \, \mathbf{V}_{\omega}^{-1} \\
  & (\gm\gn|P)_{\omega} = (\gm\gn| \frac{\text{erfc}(\omega\, r_{12})}{r_{12}} |P) \\
  & (P|Q)_{\omega} = (P| \frac{\text{erfc}(\omega\, r_{12})}{r_{12}} |Q) \text{,}
\end{align}
which, essentially, removes the long-range tail of $r_{12}^{-1}$.
If the damping frequency $\omega$ is chosen appropriately,
the sparsity of the overlap metric $S_{PQ} = \int\int \chi_P(\bfr_1) \chi_Q(\bfr_2) d\bfr_1 d\bfr_2$ 
is combined with the accuracy of the Coulomb metric $\bfV_{PQ}$.
Benchmark calculations with CDD-dRPA showed that $\omega=0.1$ introduces only mHartree deviations
from the Coulomb metric results while speeding-up the calculation by a factor 10.\cite{Luenser2017}

Our algorithm for computing the SO-CDD-MP2 energies is given in Fig.\ \ref{alg:cdd-mp2}.
First of all, $\mathbf{\po}$ and $\mathbf{\pv}$ 
are computed either from quaternion MOs or complex spinors.
The first are obtained by diagonalizing the quaternion Fock matrix in terms of 
quaternion algebra;\cite{Saue1997,Saue1999,Shiozaki2017}
the latter are solution of the complex Fock matrix (\textit{vide infra})
which is usually more efficient as highly tuned linear algebra routines
can be employed.\cite{Armbruster2008}
Then, $\mathbf{\lo}$ and $\mathbf{\lv}$ are computed by a naive
pivoted quaternion Cholesky decomposition.
The outer loop for the integral transformation step runs over
shells of auxiliary basis function rather than individual 
spherical Harmonic components.
Also the screening procedure incorporated in the sparse
matrix multiplication accounts for the degeneracy of auxiliary
basis function shells to maintain rotational invariance of the MP2 energies.
After the transformation of the AO integrals with the CDDs, $\mathbf{I}_{ai}^P$
is resorted such that the auxiliary index is the leading index.
Note that for large molecules each super block $[ai]$ usually has a different number 
of auxiliary basis functions that is much smaller than the total number
of unscreened auxiliary basis functions of the molecule.
To compute the B intermediates, the resorted integrals are transformed with the Cholesky
factors of the intermediate $\tilde{\mathbf{V}}$.
The Cholesky factorization has to be performed for each super block $[ai]$
as the resorted integrals have a different number of block specific auxiliary
basis functions $P_i$.
The Z intermediate in a selected auxiliary basis $P_i$
is computed via a symmetric rank update of the B intermediate
and than added to the Z-intermediate in the full auxiliary basis.
The dot product of the Z intermediate gives the Coulomb MP2 energy for each
quadrature point $z$.
The algorithm for the exchange part of the MP2 energy is similar to
conventional RI-MP2 implementations\cite{Feyereisen1993,Weigend1997} though
in our implementation the two outer loops run over virtual rather than occupied blocks
for a more efficient parallelization and in order to keep all occupied
block associated to at least a single virtual block in memory.
Moreover, we exploit that $E_K$ is invariant when permuting
either the two occupied or the two virtual block indices.

If only the Coulomb MP2 energy is required, as in SOS-MP2\cite{Jung2004} 
or dRPA\cite{Eshuis2010,Schurkus2016,Luenser2017},
it is more efficient to transform the square of $\mathbf{I}_{ai}^P$ only once with $\tilde{\mathbf{V}}$
for every quadrature point.
However, the time-determining step of our CDD-MP2 implementation is the exchange
energy computation.
By working with B intermediates rather than transformed integrals $\mathbf{I}_{ai}^P$
we can avoid transformations with $\tilde{\mathbf{V}}$ in the most inner loop
for the algorithm that computes $E_K$.

\section{Computational details} \label{compdet}
Our 1C- and 2C-CDD-MP2 implementation is integrated into a development
version of the DIRAC program package\cite{DIRAC17} for relativistic calculations.
The uncontracted Cartesian integrals for real large-component basis functions
were computed with the InteRest library.\cite{Repisky2013}
The exponents and weights of the numerical quadrature in Eq. \eqref{laplace}
were obtained from the minimax algorithm\cite{Takatsuka2008,Helmich-Paris2016d}
that is available as public open-source library.\cite{laplace-minimax}

All 1C and 2C Hartree--Fock calculations were performed with
Turbomole 7.2\cite{TURBOMOLE72,Furche2014}.
For the 1C and 2C calculations, we used the dscf\cite{Haeser1989} 
and ridft\cite{Haeser1989,VonArnim1998,Ahlrichs2004,Armbruster2008} module, respectively.
The relativistic SO calculations with ridft were performed with DLU approximation\cite{Peng2012,Peng2013}
to the exact 2C core Hamiltonian.\cite{Ilias2005,Kutzelnigg2005,Kutzelnigg2006}
For all HF and MP2 calculations, we employed the cc-pVTZ orbital\cite{Dunning1989} and auxiliary basis set\cite{Weigend2002} for H and the 2p elements.
For the 1C-ECP calculations, we used the cc-pVTZ-PP orbital\cite{Peterson2003} and auxiliary basis set\cite{Haettig2012} in combination with an ECP that
puts 28 electrons in the core.
For the all-electron X2C calculations, we used the Dyall valence triple $\zeta$
orbital basis set for Te.\cite{Dyall2006}
The corresponding auxiliary basis set was automatically generated by the AutoAux module\cite{Stoychev2017} of ORCA\cite{Neese2012} and provided
as supplementary material.\emph{reference to be inserted in final version}

We employed the frozen-core approximation for the following atoms:
the 1s\hoch{2} electrons of C, N, and O;
the 4s\hoch{2}4p\hoch{6} electrons of Te in ECP calculations;
the [Ar]3d\hoch{10}4s\hoch{2}4p\hoch{6} electrons of Te in all-electrons calculations.
The frozen-core approximation for valence property calculations
was presumed when designing the basis sets that we have used.

The Te-PEG-n oligomers were optimized with the Turbomole 7.2\cite{TURBOMOLE72,Furche2014} using 
the PBE density functional\cite{Perdew1996,Treutler1995} with 
D3 dispersion correction\cite{Grimme2006,Grimme2010},
and the def2-SVP orbital and auxiliary basis set\cite{Schaefer1992,Weigend2005,Weigend2006} 
in combination with an ECP with 28 core electrons.\cite{Peterson2003}
Likewise, we have optimized the structure of the Te-containing supra-molecular complex.
All structures are available as supplementary material.\emph{reference to be inserted in final version}

Unless otherwise noted, we used a target block length of 32 and
a conservative screening threshold $T_{\text{sparse}}=10^{-8}$ for the CDD-MP2 calculations.
Furthermore, we have used for the larger molecule calculations 
10 quadrature points for the numeric integration
and present results for quadrature point number 5.
The frequency for the Coulomb-attenuated integrals was set to 0.1.

\section{Results and discussion} \label{results}

\subsection{Error analysis for small molecules}

\begin{figure}
 \centering
 \begin{subfigure}[c]{0.48\textwidth}
  \centering
  \includegraphics[width=0.6\textwidth]{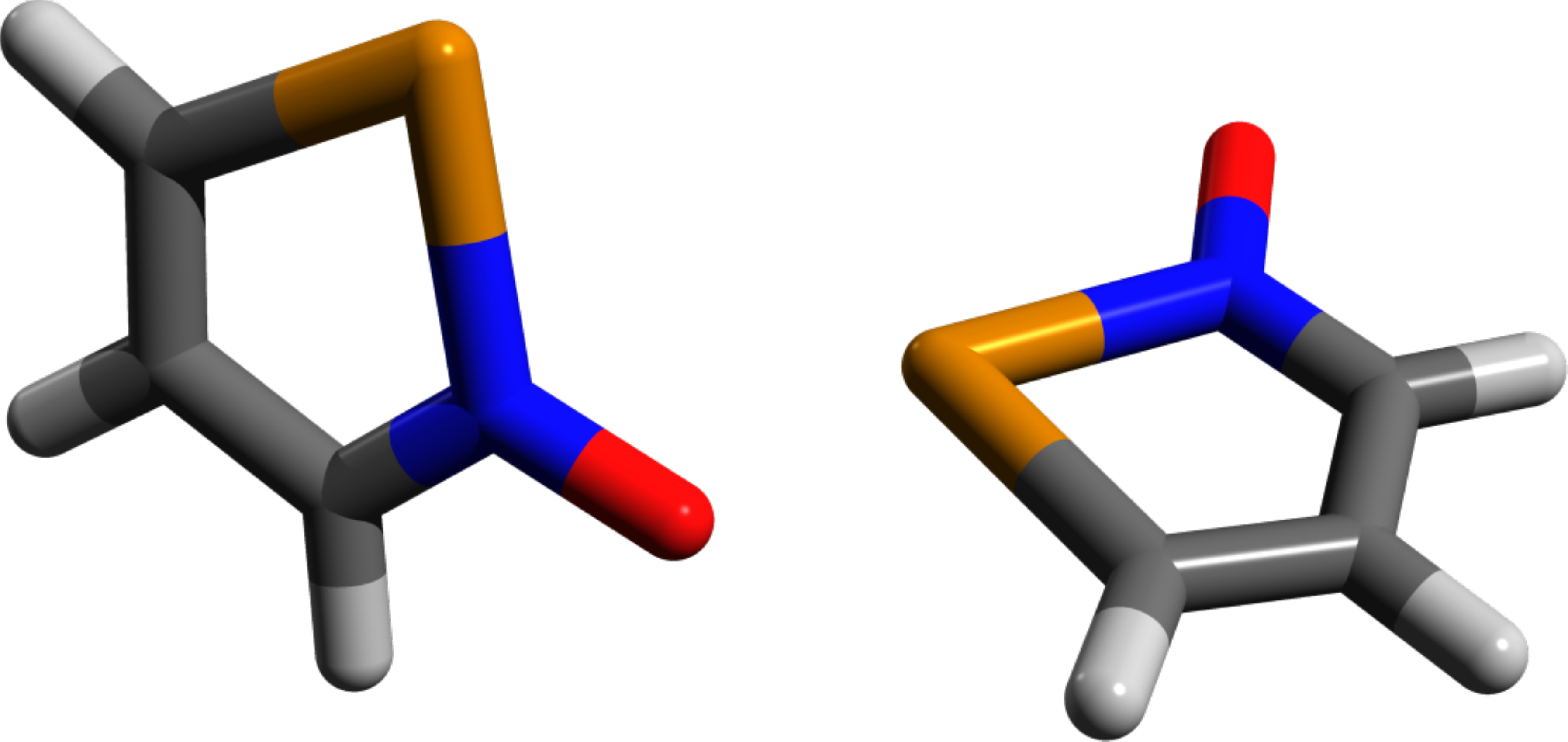}
  \caption{}
  \label{fig:small-complex}
 \end{subfigure}
\hfill
 \begin{subfigure}[c]{0.48\textwidth}
  \centering
  \includegraphics[width=.99\textwidth]{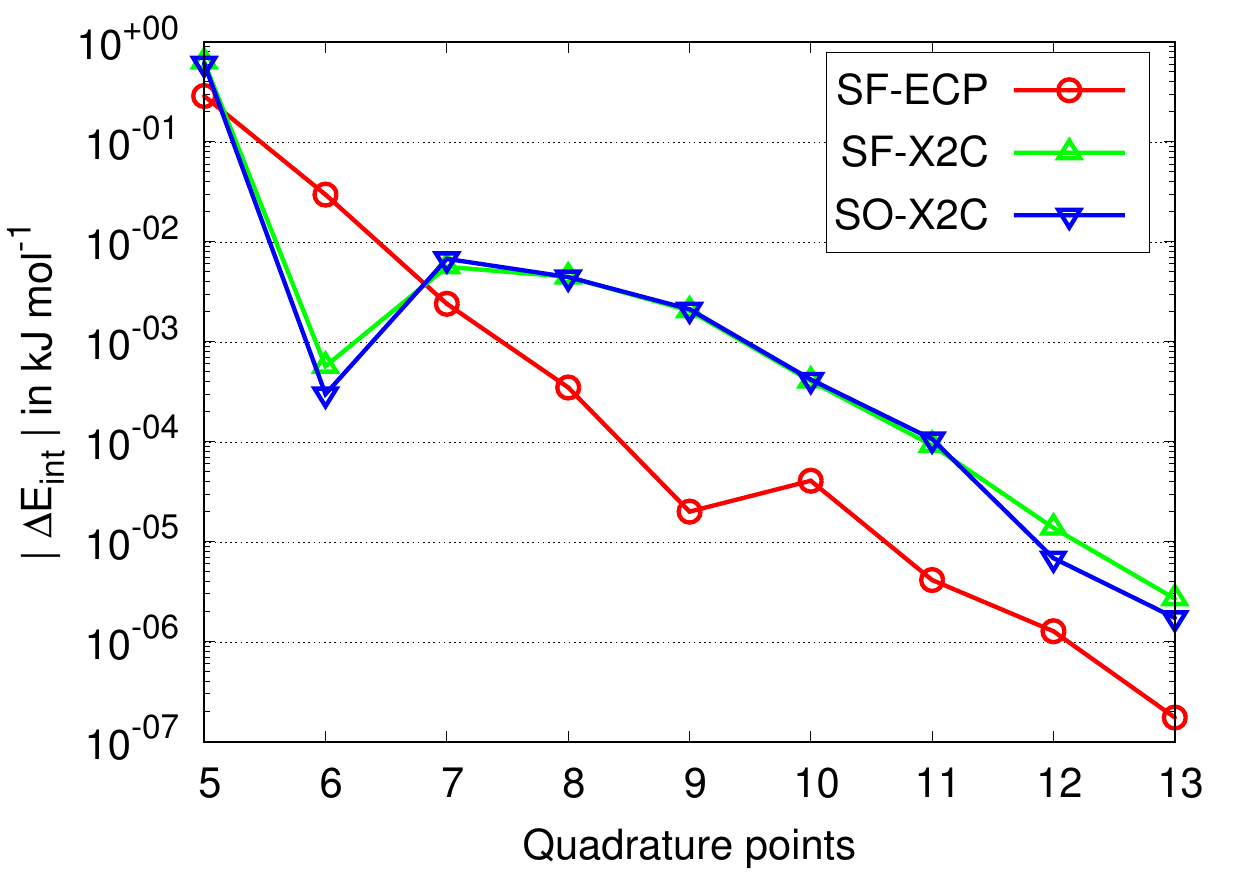}
  \caption{}
  \label{fig:lap-conv}
 \end{subfigure}
 \caption{
  Convergence of the interaction energy errors of complex (\subref{fig:small-complex}) 
  with respect to the number of quadrature points for numerical integration of the Laplace
  transform (\subref{fig:lap-conv}).
  The fitting interval ratio for the minimax algorithm for the monomer (M) and dimer (D)
  are 1C-ECP: 498. (M), 661. (D); SF-X2C: 764. (M), 1620. (D); SO-X2C: 764. (M) and 1635. (D).
}
 \label{fig:small-dimer}
\end{figure}

The numerical errors of MP2 correlation energies introduced by our CDD implementation
are caused by (1) the RI approximation, (2) the numerical integration of the Laplace
transform, and (3) the neglect of blocks with a small norm in sparse intermediates.
For calculations of small molecules with sufficiently accurate sparse-matrix thresholds $T_{\text{sparse}}$,
nearly all blocks are kept with the pursued screening protocol.
Thus, when investigating small molecules we focus on the errors that are caused by the RI approximation and the numerical
integration only and set $T_{\text{sparse}}$ to zero.
For a supramolecular complex of two tellurazol oxide monomers
the convergence of the MP2 interaction with respect to number of quadrature points
is shown in Fig.\ \ref{fig:small-dimer} for the 1C-ECP, SF-X2C, and SO-X2C Hamiltonians.
For all three Hamiltonians the errors in the MP2 interaction energy converge rapidly to zero.
For the all-electron calculations we have used uncontracted basis sets 
due to technical limitations of the X2C implementation in the Dirac program package.
Moreover, for those calculations all virtual orbitals with an orbital energy larger than 40~a.\ u.\ 
were frozen to facilitate the reference calculation with the conventional Dirac MP2 implementation\cite{Laerdahl1997}.
All-electron calculations usually feature a much larger ratio of the maximum to minimum
orbital energy denominator ($R=\max(x)/\min(x)$) as given in the caption of Fig.\ \ref{fig:small-dimer},
which requires more quadrature points to reach the same accuracy by the minimax algorithm.\cite{Takatsuka2008,Helmich-Paris2016d}
This is exactly what can be observed in  Fig.\ \ref{fig:small-dimer}, though
for the present example, only one or two additional quadrature points are sufficient
to reach the same accuracy in the X2C and ECP calculation.

The HF and correlation energy contribution to the MP2 interaction energies of the two Tellurazol oxide monomers are compiled
in Tab.\ \ref{tab:small-complex}.
Using the RI approximation leads to a slight overestimation of the interaction energy.
Due to a relatively large automatically generated auxilliary basis set for Tellurium 
when employing the X2C Hamiltonian, the RI error of all-electron calculations is significantly 
smaller than the one of the ECP calculations.
Nevertheless, the RI errors are satisfactory if one considers the inherent 
methodological error of the MP2 method. 
Eventually, the RI errors would also decrease when using larger
auxiliary (and orbital) basis sets. 

\subsection{Performance for large linear molecules}

\begin{figure}
 \centering
 \includegraphics[]{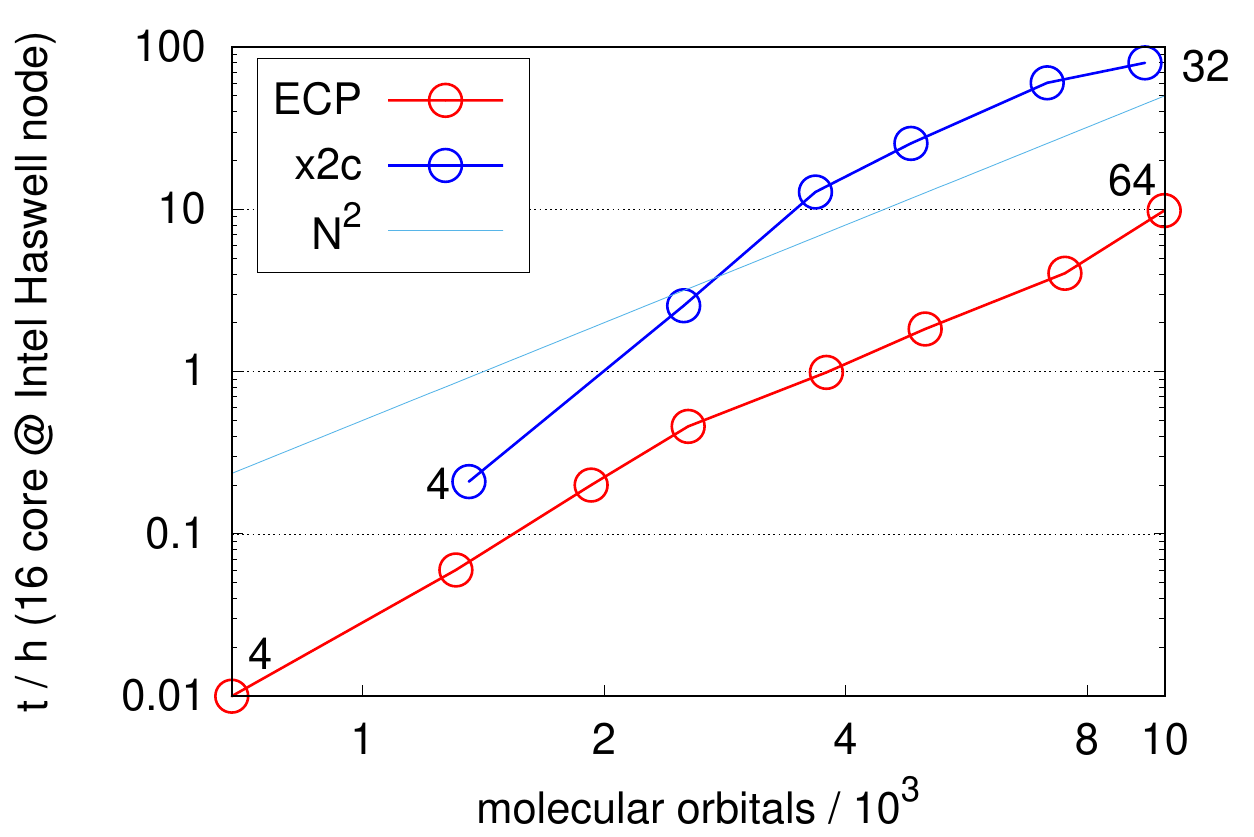}
 \caption{
  Wall time of 1C-ECP and X2C CDD-MP2 calculations on Te-PEG-n with $\text{n}=\{4,8,12,16,24,32,48,64\}$
  for a single quadrature point on a single Intel Haswell node with 16 threads.
}
 \label{fig:timings}
\end{figure}

We investigated the scaling with the system size  (Fig.\ \ref{fig:timings}) 
of 1C and 2C CDD-MP2 for linear chains of Tellurium-substituted poly-ethylene glycol
oligomers Te-PEG-n with  $\text{n}=\{4,8,12,16,24,32,48,64\}$. 
All calculations were performed in parallel with 16 threads on a Intel Haswell node.
The largest 1C-ECP CDD-MP2 calculation (n=64) involved 9987 and 27208 spherical Harmonic orbital and auxiliary
basis functions, respectively.
For the largest X2C CDD-MP2 calculation (n=32) 9449 spherical Harmonic orbital and 41523
auxiliary basis functions were employed.
For a single quadrature point, the largest 1C-ECP CDD-MP2 calculation (n=64) took approximately 10 hours;
the largest X2C CDD-MP2 calculation (n=32) 3 days and 8 hours.
The contraction to exchange energy is for the 1C-ECP and X2C
calculation by far the most time-consuming step and takes approximately 57 and 96 \%
of the total run time, respectively.
This is expected as the $E_K$ contraction is the only computational
step with a formal $\mathcal{O}(N^5)$ scaling if no blocks can be screened.
Additionally, the $E_K$ contraction has the largest formal spin-orbit overhead (28)
of all computational steps.
Therefore, almost the total computation time is spent on the 2C exchange contraction.
Improvements on the performance of the exchange contraction could be attained
by different integral decomposition techniques like the tensor hyper-contraction\cite{Hohenstein2012}.
Those can eventually result in an $\mathcal{O}(N^4)$-scaling implementation
that offers the possibility to avoid the quaternion commutator term.

\begin{figure}
 \centering
 \includegraphics[]{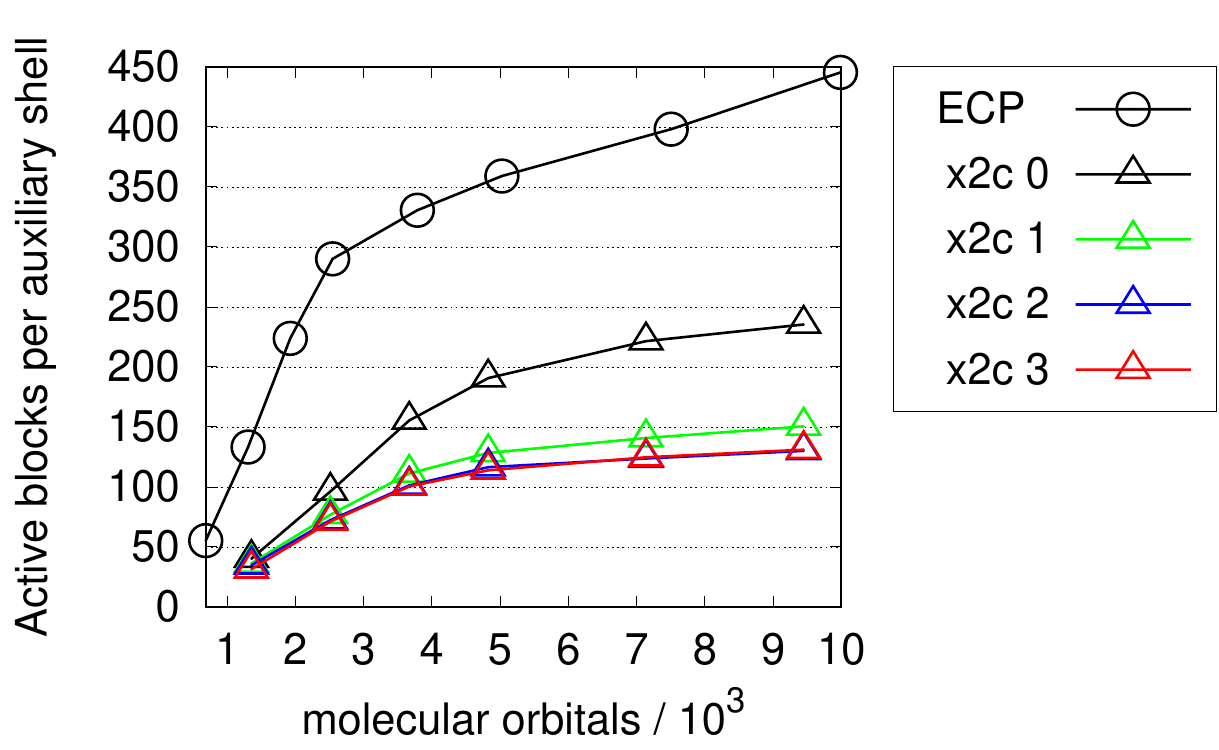}
 \caption{
  Average number of active blocks per auxiliary basis function shell for 1C-ECP and X2C calculations on Te-PEG-n with $\text{n}=\{4,8,12,16,24,32,48,64\}$.
}
 \label{fig:blocks}
\end{figure}

For the timings of the 1C-ECP and X2C CDD-MP2 calculation we observe an effective quadratic
scaling with respect to the number spherical Harmonic orbital basis functions --- a measure
of the size of a molecule.
We expect a better scaling if we would run the calculations in serial as I/O
of transformed three-index integrals and B intermediates has also
a significant contribution to the timings, which is at the moment not parallelized by
our shared-memory OpenMP parallelization.
Eventually, linear scaling of nearly all computational resources should be observable
since the average number of significant, unscreened sparse CDD blocks [ai] per auxiliary shell converges
to a constant value in the asymptotic limit (Fig.\ \ref{fig:blocks}).
That number is for the 1C-ECP calculation roughly a factor of two smaller
than for each quaternion unit in the X2C calculations.
Compared to the 1C-ECP calculations, we need more orbital basis functions 
for the all-electron calculations
and would expect a larger number of active blocks.
However, we observe the opposite.
It is the large number of steep s, p, and d short-range AOs in both the orbital and auxiliary Te basis set
that leads to a much smaller number of CDD blocks [ai] for each quaternion unit.

\begin{figure}
 \centering
 \includegraphics[]{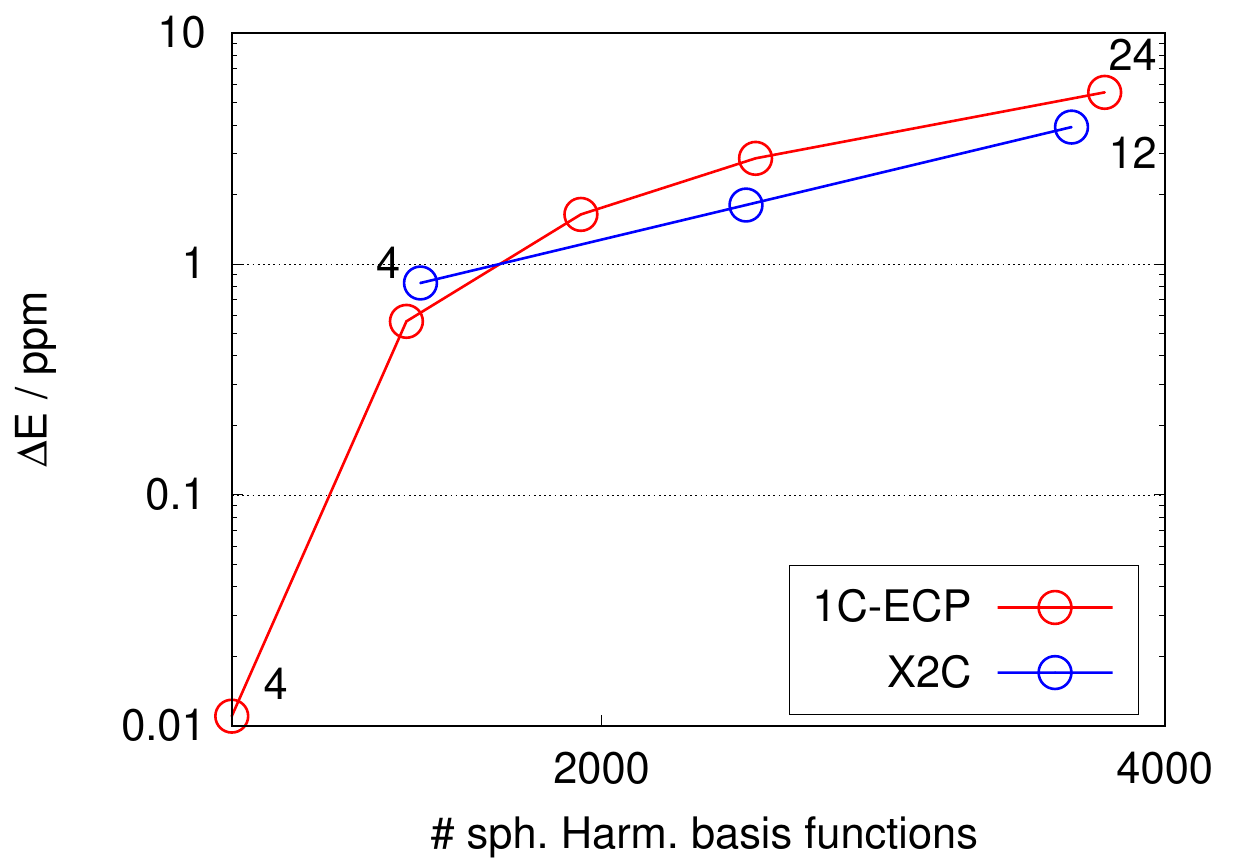}
 \caption{Relative errors in ppm for 1C-ECP and X2C calculations on Te-PEG-n with $\text{n}=\{4,8,12,16,24,32,48,64\}$.}
 \label{fig:errors}
\end{figure}

The largest impact on the run time and accuracy of our CDD-MP2 implementation is the
screening of sparse matrices. We have chosen a rather conservative sparse-matrix screening
threshold of $T_{\text{sparse}}=10^{-8}$. The relative errors in ppm for the linear Te-PEG-n oligomers
are given in Fig.\ \ref{fig:errors} and are always smaller than 10 ppm.
Compared to the error introduced by the RI approximation,
the truncation error is negligible.
Note that we were not able to calculate the larger oligomers without screening due to
the wall time limit on the compute cluster and limited hard disk size.
The errors are increasing with the system size and converge to a finite value when increasing 
the system size.
We note that looser screening threshold of $T_{\text{sparse}}=10^{-5}$ and $10^{-6}$
lead to unacceptably large errors that can be attributed to the transformation with
the inverse or inverse square root of the Coulomb metric $V_{PQ}$.
A more compact local RI basis should allow for calculations with looser thresholds,
which is at the moment not available.

\subsection{Performance for large bulky molecules}

\begin{figure}
 \centering
 \begin{subfigure}[c]{0.48\textwidth}
  \includegraphics[width=0.7\textwidth]{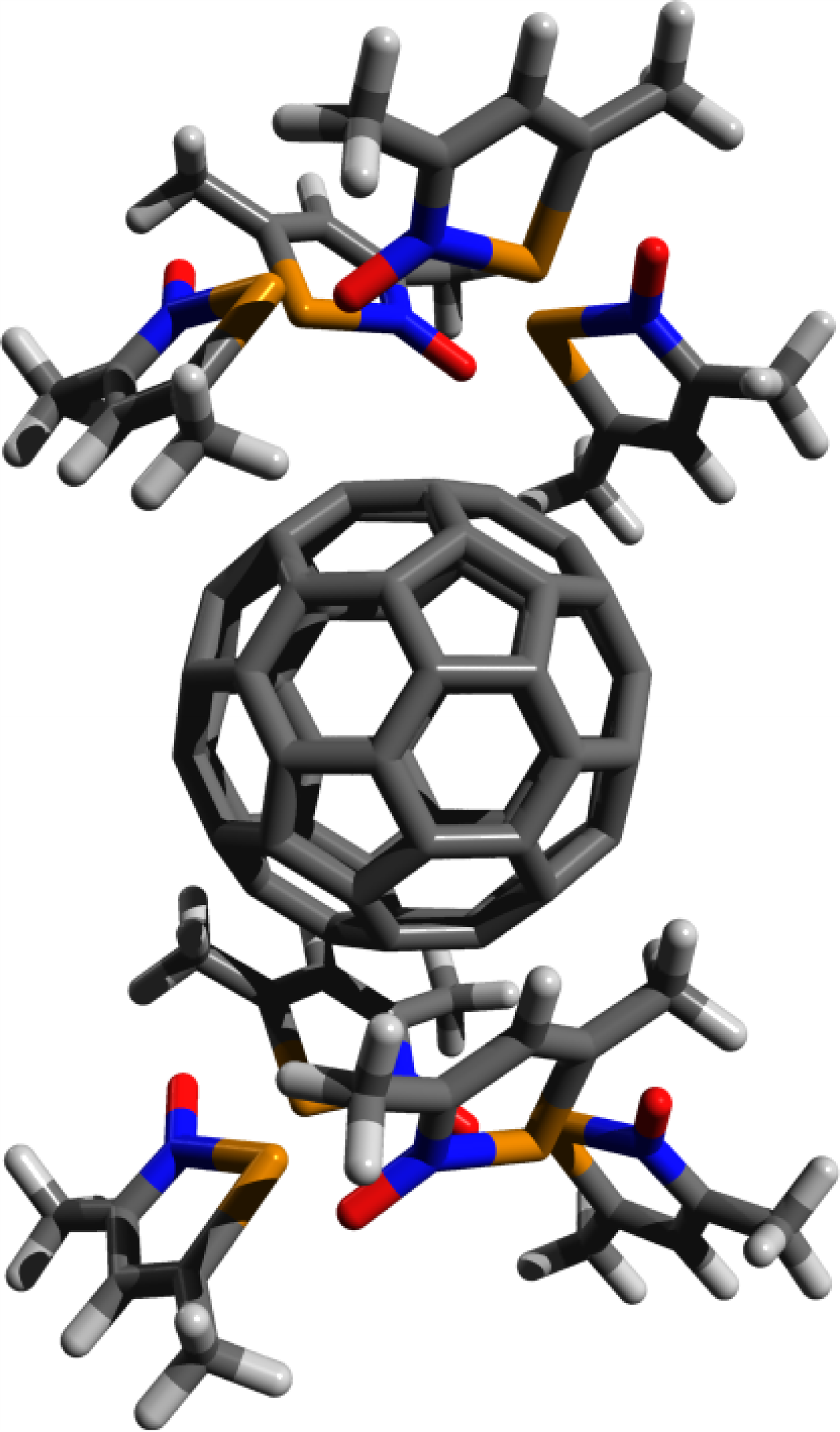}
  \caption{}
  \label{fig:complex}
 \end{subfigure}
\hfill
 \begin{subfigure}[c]{0.48\textwidth}
  \scalebox{0.7}{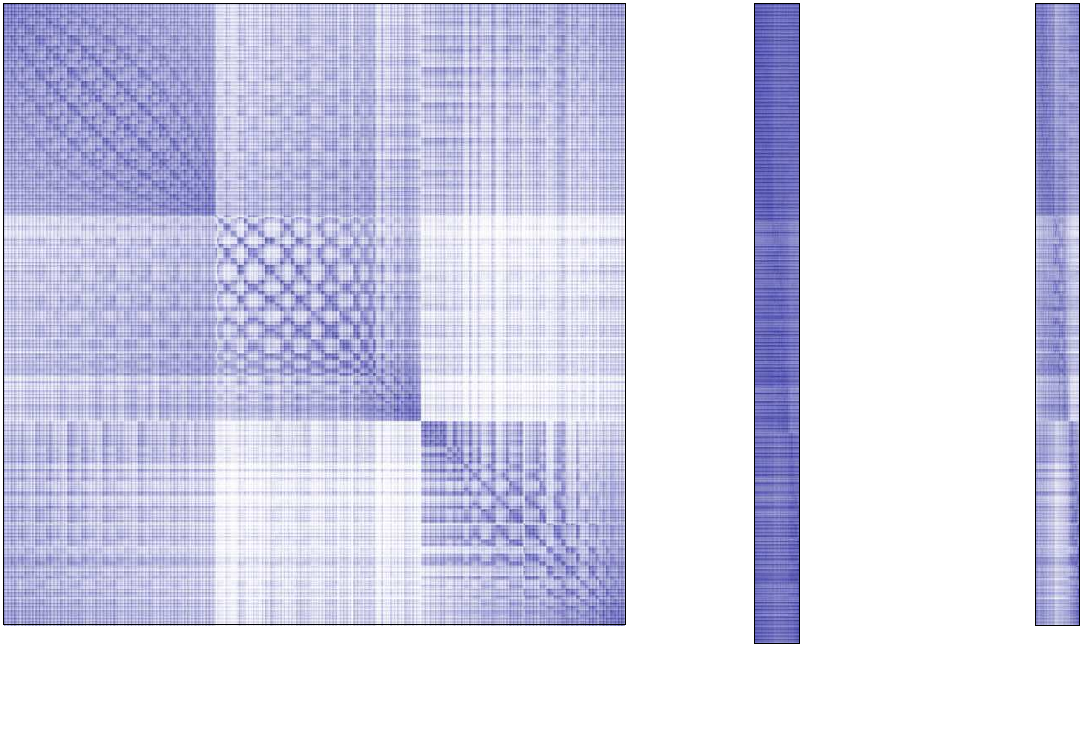}
  \caption{}
  \label{fig:dens}
 \end{subfigure}
  \caption{
   A Te-containing supramolecular complex (\subref{fig:complex})
   with non-relativistic occupied 
   pseudo-density matrix, pivoted Cholesky factors, and localized occupied molecular orbitals
   (\subref{fig:dens}).
  }
 \label{fig:complex-den-cdd}
\end{figure}

We also investigated the performance of our CDD-MP2 implementation for 
a bulky molecule, i.e. a supramolecular complex of the Buckyball C\tief{60}
bound to two four-membered rings that are closed by strong Te-O non-covalent bonds
(Fig.\ \ref{fig:complex}).\cite{Ho2016}
Computing inter-molecular interaction energies of supramolecular complexes 
with heavy elements are potential applications of our 2C CDD-MP2 and we would like
to study the feasibility of such calculations.

For the 1C-ECP calculation, we show in Fig.\ \ref{fig:dens}
contour plots of occupied pseudo-density matrix, its pivoted Cholesky
factor, and localized occupied molecular orbitals obtained 
from the Foster-Boys localization scheme.\cite{Foster1960}
In contrast to the larger linear Te-PEG-n chains (\textit{vide supra}),
both the occupied pseudo-density matrix and 
pivoted Cholesky factor (CDD) are dense and, thus, not suited for our block-wise screening
procedure.
We have also pursued a localization of the occupied and virtual
molecular orbitals (LMO) by the Foster-Boys procedure that minimizes the
orbital variance.\cite{Foster1960}
As can be seen from Fig.\ \ref{fig:dens}
those localized molecular orbitals can be much more compact
than the CDD.
Nevertheless, the current screening scheme based on the inspection of CDD / LMO blocks
did not lead to any negligible contributions for such bulky molecule.
This is not surprising if one considers the diameter of the molecular complex 
in Fig.\ \ref{fig:complex} of about 21.1~{\AA}.
The 1C CDD-MP2/cc-pVTZ-PP calculation of that Te-containing complex required
5288 and 14056 orbital and auxiliary basis functions, respectively.
We needed a similar number of basis functions (5027/13704)
for the 1C-ECP CDD-MP2 calculation of Te-PEG-32. 
The length of that linear molecule is 148.7~{\AA}.
Such one-dimensional systems offer a much better possibility for as screening
as whole CDD/LMO blocks can be easily discarded.
A similar screening efficiency could also be acheived for bulky molecules
if they have a diameter of e.g.\ 100~{\AA} and more but this would exceed
computing resources that are generally available.

At the moment, an obvious direction towards an improved performance for bulky
molecules along present lines seems to be unclear.
The orbital spread of LMOs could be improved by minimizing
higher orders of the orbital variance.\cite{Jansik2011,Hoeyvik2012,Hoeyvik2012a}
One could also consider to return to an AO-basis implementation 
that screens shell pairs based on (distance-dependent) pseudo Schwarz estimates.\cite{Haeser1993,Lambrecht2005a,Maurer2012}
A combination with the RI approximation or an multipole expansion
of the far-field Coulomb interaction\cite{White1994} seem to be yet unexplored.

\section{Conclusions}
In the present article, we have introduced the relativistic Cholesky-decomposed
density (CDD) matrix MP2.
The working equations are formulated in terms of the usual intermediates
of RI-MP2 and are obtained by substituting the occupied and virtual quaternion pseudo-density matrices
of our previously proposed 2C AO-MP2 by the corresponding pivoted quaternion
Cholesky factors.
While working within the Kramers-restricted formalism, we obtain a formal spin-orbit overhead of 16 and 28
for the Coulomb and exchange contribution to the 2C MP2 correlation energy, respectively,
compared to a non-relativistic spin-free CDD-MP2 implementation.
This reduced spin-orbit overhead is a consequence of the quaternion algebra
which could also be exploited for
any other conventional or approximate algorithm for 2C MP2 energies.
The errors that were introduced by the RI approximation and the numerical integration
were investigated for a small Te-containing supramolecular complex
and are negligible if the inherent methodological error of MP2 as well as
the basis set incompleteness error are considered.
The quaternion Cholesky factors become sparse for large linear systems and,
by adapting the block-wise screening, block sparse-matrix multiplication
algorithm of Ochsenfeld and co-workers,\cite{Luenser2017} we were able to compute
1C-ECP MP2 correlation energies for a linear Te-containing polyethylene glycol chain
Te-PEG-64 with more than 400 atoms and roughly 10,000 orbital basis functions within 10 h by using
16 threads.
The X2C all-electron MP2 calculation of the half-size chain (Te-PEG-32) needed roughly the same number
of basis functions, but due to the spin-orbit overhead much longer 3 d and 8 h with 16 threads.
The total run time for both 1C and 2C calculations
was dominated by the contraction to the exchange energy.
This computational step has still the original MP2 $\mathcal{O}(N^5)$ scaling 
if no blocks can be screened.
For the linear chains we observed an effective quadratic scaling of the total
wall time with the system size.
We have also investigated a bulky Te-containing supramolecular complex.
For such bulky, three-dimensionally extended molecules the present implementation
is unfortunately less suited as the CDDs are dense.
Improvements on the performance of 2C-MP2 energies calculation for large and bulky molecular systems
will be investigated in the near future.

\section{Acknowledgments}
B.\ H.-P.\ acknowledges gratefully financial support from the German Research Foundation DFG
(Grant No.\ HE 7427/1-1) and from the Netherlands Organisation 
for Scientific Research NWO by a Veni fellowship (Grant No.\ 722.016.011).
M.\ R.\ acknowledges financial support by the Research Council of Norway through its
Centres of Excellence scheme, project number 262695. 
Computer time at the Dutch national super computer Cartesius granted by the NWO
is very much appreciated.
B.\ H.-P.\  would like to thank Georgi L.\ Stoychev for discussions on automatized 
auxiliary basis sets and Florian Weigend and Uwe Huniar for providing generous
support on 2C Hartree-Fock calculations with Turbomole.

\section{Supporting information}
The corresponding auxiliary basis set of the Dyall valence triple $\zeta$
basis for Tellurium was obtained by an automated fitting procedure (AutoAux)
that is available in the ORCA quantum chemistry package and is provided
as supplementary material.
Furthermore, all Cartesian coordinates of molecular structure used
in the present article are made available.


\providecommand{\latin}[1]{#1}
\providecommand*\mcitethebibliography{\thebibliography}
\csname @ifundefined\endcsname{endmcitethebibliography}
  {\let\endmcitethebibliography\endthebibliography}{}

\newpage

\begin{figure}
\begin{algorithmic}[]
\State compute pseudo-density matrices $\mathbf{\po}$ and $\mathbf{\pv}$ and  BCSR packing
\State CDD $\underline{\mathbf{L}}$ and $(\bar{\mathbf{L}})^H$ and BCSR packing
\State Schwarz estimates $Q_{\gm \gn,\omega} = \sqrt{(\gm\gn|\gm\gn)_{\omega}}$ and BCSR packing
\State \# integral transformation
\ForAll{auxiliary shells}
 \State compute all $(\gm\gn|P)_{\omega} \, \forall \, \gm,\gn : Q_{\gm \gn} \ge T_{\text{Sparse}}$
 \ForAll{z}
 	\State BCSR quaternion multiplication: $\mathbf{I}_{\gm i}^P = (\gm\gn|P)_{\omega} \times \underline{\mathbf{L}}_{\gn i}$
 	\State BCSR quaternion multiplication: $\mathbf{I}_{a i}^P =  \bar{\mathbf{L}}_{a \gm}^* \times \mathbf{I}_{\gm i}^P$
 	\State write sparse matrices $\mathbf{I}_{a i}^P \, \forall P \in \text{aux. shell}$ to disk
 \EndFor
\EndFor
\State resort $\mathbf{I}_{ai}^P$ and make $P$ leading index
\State compute $\tilde{\mathbf{V}} = \mathbf{V}_{\omega}^{-1}\, \mathbf{V} \, \mathbf{V}_{\omega}^{-1}$
\ForAll{z}
 \ForAll{ super blocks [ai] }
   \ForAll{q}
    \State read $\mathbf{I}_{ai}^P$
    \State get subset of $\tilde{\mathbf{V}}_{P_i Q_i}$ for subset of aux. BF $P_i$ part of [ai]
    \State Cholesky decomposition: $\tilde{\mathbf{V}}_{P_i Q_i} = L_{P_i,\tilde{P}_i} L_{Q_i,\tilde{P}_i}$
    \State $B_{ai}^{\tilde{P}_i,q} = L_{P_i,\tilde{P}_i} I_{ai}^{P_i,q}$
    \State $Z_{\tilde{P}_i\tilde{Q}_i} = B_{ai}^{\tilde{P}_i,q} B_{ai}^{\tilde{Q}_i,q}$
    \State $Z_{PQ} +=\  Z_{\tilde{P}_i\tilde{Q}_i}$
  \EndFor
 \EndFor
 $e_J^{(z)} +=  \text{tr}[\mathbf{Z}^T \mathbf{Z}]$
\EndFor
\ForAll{z}
 \ForAll{batch of [a]}
  \State read $\mathbf{B}_{a*}^P$
  \ForAll{batches of [b]}
   \State read $\mathbf{B}_{b*}^P$
   \ForAll {$[a]$}
    \ForAll {$[b] \le [a]$}
     \ForAll {$[i]$}
      \ForAll {$[j] \le [i]$}
       \State find common set of aux. BF for $[ai]$ and $[bj]$
       \State $e_K^{(z)} += B_{ai}^P B_{bj}^P B_{aj}^Q B_{bi}^Q$ according to Eq.\ \eqref{cdd-ek}
      \EndFor
     \EndFor     
    \EndFor
   \EndFor
  \EndFor
 \EndFor
\EndFor
\end{algorithmic}
\caption{CDD-MP2 algorithm}
\label{alg:cdd-mp2}
\end{figure}

\newpage

\begin{table}
 \caption{Interaction energies of two tellurazol oxide molecules given in kJ mol\hoch{-1}.}
 \label{tab:small-complex}
 \begin{tabular}{lp{2.5cm}p{2.5cm}p{2.5cm}p{2.5cm}}
 \multicolumn{1}{c}{Hamiltonian} &
 \multicolumn{1}{c}{HF} &
 \multicolumn{1}{c}{RI-MP2 (corr.)} &
 \multicolumn{1}{c}{MP2 (corr.)} &
 \multicolumn{1}{c}{$\Delta$ RI} \\[0.3em] \hline \\[-0.8em]
%
%
  1C-ECP      & -62.6831 &  -8.3451 &  -8.2763 & -0.0688 \\
  SF-X2C      & -61.6723 &  -9.9347 &  -9.9223 & -0.0124 \\
  SO-X2C      & -60.2273 & -10.2285 & -10.2205 & -0.0080 \\[0.5em]
%
\hline
 \end{tabular}
\end{table}

\end{document}